\documentclass[epj]{svjour}
\usepackage{latexsym}
\begin{document}

\title{ The radial Teukolsky equation for Kerr-Newman-de Sitter geometry: Revisited  }

\author{M. Horta\c{c}su} 

\institute{ Department of Physics, Mimar Sinan G\"{u}zel Sanatlar \"{U}niversitesi, Istanbul, Turkey }
\mail{hortacsu@itu.edu.tr}
\date{Received: date / Revised version: date}

\abstract{ We use Heun type solutions given in \cite{Suzuki} for the radial Teukolsky equation, written in the background metric of the Kerr-Newman-de Sitter geometry, to calculate the quasinormal frequencies for polynomial solutions and the reflection
coefficient for waves coming from the de Sitter horizon and reflected at the outer horizon of the black hole.}
\maketitle
 
\section{Introduction}
\label{intro}
It is well known that the wave equation for a scalar particle, in the background of any D type metric, gives Heun type solutions \cite{Batic}. A recent paper also showed the separability of conformally coupled scalar field equation in general (off-shell) Kerr-NUT-AdS spacetimes in all dimensions \cite{Finnian}. Work in this direction was done in the past. To mention one leading paper, one may cite Carter \cite{Carter}, who showed that the scalar wave equation is separable in the Kerr-Newman-de Sitter geometries. Teukolsky \cite{Teukolsky} generalized this result for spinors, electromagnetic fields, gravitational fields and gravitinoes in the Kerr-Newman and Kerr-Newman-de Sitter class of geometries. In this respect, one can also give \cite{Chen,Benone,Cebeci,Jiang,Vieira}. 

In two papers \cite{Suzuki,Suzuki1}, Suzuki et al. obtained the exact solutions of the Teukolsky equations in terms of Heun type functions \cite{Heun,Ronveaux,Slavyanov,Hortacsu,Tolga3}. Suzuki et al. expanded the Heun solutions in terms of infinite series of hypergeometric functions, and used this infinite series solution in their work. Quoting from their second paper \cite{Suzuki1}, ''they chose the solution which satisfied the incoming boundary conditions at the outer horizon of the black hole and examined the asymptotic behavior at the de Sitter horizon. They also evaluated the absorption rate of the Kerr-de Sitter and the Kerr-Newman-de Sitter black holes by using the analytic solution expanded in terms of an infinite series of hypergeometric functions \cite{Suzuki1}. They constructed the conserved current by evaluating the Wronskian, and obtained an expression of the absorption rate. From this, they showed explicitly that super-radiance occurs for the boson, similarly to the Kerr geometry case \cite{Mano}. Then, they derived an analytic expressions of the incident, the reflection and the transmission amplitudes. They also derived the conserved current from which they derived the absorption rate in terms of an infinite series of hypergeometric functions. They also studied the asymptotic limits of their solutions."

A peculiar characteristic of the Teukolsky equations is that, after one obtains the wave equations for the radial and angular variables, one finds that they are in very similar forms. In fact, sometimes they can be made exactly the same, by choosing the correct transformation, as in the Kerr-de Sitter case \cite{Gibbons,Tolga2}, in the limit when the mass of the black hole goes to zero.

Heun functions were not very popular even at the end of the twentieth century. In the last decade, in different Citation Indices, one finds that the number of papers using Heun function solution \cite{Heun,Ronveaux,Slavyanov,Hortacsu,Tolga3} more than doubled. In WOS or Scopus, one can find many papers which give their results in terms of these functions. One can also find connection formulae for Heun functions expanded around different points \cite{Dekar,Hortacsu1}. Furthermore, new mathematical software, like Maple, includes Heun solutions. 

In many examples, once one knows that the solution will be of the Heun form and the transformation of coordinates to obtain that solution, given by Maple, I believe, it is easier to solve the equation by hand, since, Maple often gives page long solutions. These solutions are much shorter if the calculation is done by hand.

Here we used the wave equations given in \cite{Suzuki} for the radial case, obtained the Heun solutions around different points for the 
Kerr-Newman-de Sitter metric and used the exact solutions in terms of Heun functions, instead of using infinite series expansions in terms 
of hypergeometric functions \cite{Suzuki,Suzuki1}. By using the connection formulae given in \cite{Dekar}, we calculate the reflection 
coefficients for waves coming from the de Sitter horizon in a closed form.

In the next section, we summarize  the relevant information given in \cite{Suzuki} and give the quasinormal frequencies for the polynomial solutions for this equation. In section 3, we solve the wave equation using 
expansions both for around the outer horizon and around the inner de Sitter horizon exactly in terms of General Heun functions. Before doing this, we use a transformation to bring infinity to unity, and the outer horizon to zero. Then, we use these solutions and the 
connection formulae to write the reflection coefficient at the outer horizon. We end by our conclusions.

\section{The radial Teukolsky equation for Kerr-Newman-de Sitter geometry}
\label{sec:1}

Suzuki et al. \cite {Suzuki} gives the radial Teukolsky equation as 
\begin{eqnarray}
&&
\Delta^{-s} \frac{d}{dr} \Delta^{s+1}\frac{dR}{dr}+ 
\frac{1}{\Delta} 
\Big((1+\alpha)^{2} \big(K-\frac{eQr}{1+\alpha}\big)^{2}
{}\nonumber\\
&&
-is(1+\alpha)\big( K-\frac{eQr}{1+\alpha}\big)\frac{d\Delta}{dr}\Big) R 
+\Big(4is(1+\alpha)\omega r
 -\frac{2\alpha}{a^{2}}(2s+1)r^{2} +2s(1-\alpha)-2iseQ-\lambda\Big) R =0,
\end{eqnarray}
with five regular singularities at $r_{+}$, $r_{-}$, $r^{d}_{+}$, $r^{d}_{+} $ and infinity. In the words of \cite{Suzuki}, $r_{+}$ is the outer horizon, $r_{-}$ is the inner horizon, $r^{d}_{+}$ maybe called the outer de Sitter horizon, $r^{d}_{-}$ may be called the inner de Sitter horizon.
Here $\Lambda$ is the cosmological constant, $\alpha=\frac{\Lambda a^{2}}{3}$, $M$ is the mass of the black hole, $\alpha M$ its angular momentum,  
$Q$ is its charge, $K=\omega(r^{2}+a^{2})$, and
\begin{equation}
\Delta = (r^{2}+a^{2})\big(1-\frac{\alpha}{a^{2}}\big)-Mr+Q^{2}   )
  = - \frac{\alpha}{a^{2}}(r-r_{+})(r-r_{-})(r-r^{d}_{+}) (r-r^{d}_{}).
\end{equation}
By using the new variable \cite{Suzuki}
\begin{equation}
 z= \Big( \frac{r_{+}-r^{d}_{-}}{r_{+}-r_{-}}\Big)\Big( \frac{r-r_{-}}{r-r^{d}_{-}} \Big),
\end{equation} 
one gets a new  wave equation. Again quoting \cite{Suzuki}, "the new equation has regular singularities at $0,1,z_{r},z_{inf}$ and at infinity." Note that now for $r$ equal 
to $r^{d}_{-}$, $z$ goes to infinity.
Furthermore,what would be the de Sitter horizon, $r^{d}_{+} $, they call $z_{r}$, where
\begin{equation}
 z_{r}= \Big( \frac{r_{+}-r^{d}_{-}}{r_{+}-r_{-}} \Big)
 \Big( \frac{r^{d}_{+}-r_{-}}{r^{d}_{+}-r^{d} _{-}} \Big),
\end{equation}
 and
\begin{equation}
 z_{inf}= \big( \frac{r_{+}-r^{d}_{-}}{r_{+}-r_{-}} \big).
 \end{equation} 
We see that both of these latter variables take negative values, outside our realm of interest. 
We take the point where $r$ equal to $r^{d}_{-}$ as the de Sitter horizon. In \cite{Suzuki1}, the authors use a different independent variable $x=1-z$, we choose to use the variables in \cite{Suzuki}.
 
As shown in \cite{Suzuki}, one can factor out the singularity at $z=z_{inf}$ using the transformations
\begin{equation}
R(z)= z^{B_{1}} (z-1)^{B_{2}}(z-z_{r})^{B_{3}}(z-z_{inf})^{2s+1} g(z).   
\end{equation}
We think this is very remarkable, since here, by a single s-homotopic transformation, i.e. by multiplying the dependent variable by a power, 
one  gets rid of both linear and quadratic powers of $\frac{1}{z-z_{inf}}$ multiplying the dependent variable, as well as the same term 
multiplying the derivative of the dependent variable. Normally, in differential equations, such a transformation gets rid of only one term, 
usually the inverse quadratic term in the original equation. Only the special form of the used metric, as well as the suitable choice of the transformation used in \cite{Suzuki} enables this important result.

Here
\begin{equation}
B_{1} = \frac{1}{2} \Big( -s \pm  
i \big( \frac{2(1+\alpha)a^{2} (\omega(r^{2}_{-} +a^{2} ) -am
-\frac{eQr_{-}}{1+\alpha})} {\alpha (r^{d}_{+}- r_{-})(r^{d}_{+}-r_{-})(r_{+}-r_{-})}-is \big) \Big),
\end{equation}
\begin{equation}
B_{2} = \frac{1}{2} \Big( -s \pm i
\big( \frac{2(1+\alpha)a^{2} (\omega(r^{2}_{+} +a^{2} ) -am
-\frac{eQr_{+}}{1+\alpha})} {\alpha (r^{d}_{+}- r_{+})(r^{d}_{+}-r_{+})(r_{-}-r_{+})}-is \big) \Big),
\end{equation}
\begin{equation}
B_{3} = \frac{1}{2} \Big( -s \pm
i \big( \frac{2(1+\alpha)a^{2} (\omega((r^{d}_{+})^2 +a^{2} ) -am
-\frac{eQr^{d}_{+}}{1+\alpha})} {\alpha (r_{-}- r^{d}_{+})(r^{d}_{+}-r_{-})(r_{+}-r^{d}_{+})}-is \big) \Big).
\end{equation}
After all these transformations are made, we end up with a wave equation for $g(z)$ which reads
\begin{equation}
\Big(\frac{d^{2}}{dz^{2}}
+\big( \frac{2B_{1}+s+1}{z}+\frac{2B_{z}+s+1}{z-1}+\frac{2B_{3}+s+1}{z-z_{r}}\big) \frac{d}{dz}
+ \frac{\sigma_{+}\sigma_{-}z+v}{z(z-1)(z-z_{r})} \Big)g(z)=0,
\end{equation}
where
\begin{equation}
 \sigma_{\pm}= B_{1}+B_{2}+B_{3}+2s+1 
 +\frac{1}{2}\Big(-s\pm i \big(-i s
 +\frac{2(1+\alpha)a^{2} (\omega((r^{d}_{-})^2 +  a^{2})-am-\frac{eQr^{d}_{-}}{1+\alpha})}{\alpha(r^{d}_{+}- r_{+})(r^{d}_{+}-r_{+})(r_{-}-
 r_{+})}\big)\Big),
\end{equation}
and 
\begin{equation}
v= A(B+C)+D+E,
\end{equation}
where
\begin{equation}
A=\frac{2a^{4}(1+\alpha)^{2} (r_{+}- r^{d}_{+})^{2}(r_{+}-r^{d}_{-})^{2}(r_{-}-r^{d}_{-}) (r^{d}_{+}-r^{d}_{-})}{\alpha^2 T( r_{+}-r_{-})},
\end{equation}

\begin{equation}
B=-\omega^{2}r^{3}(r_{+}r_{-}-2r_{+}r^{d}_{+} +r_{-}r^{d}_{+})
+2a\omega(a\omega-m)r_{-}(r_{+}r^{d}_{+}-r^{2}_{-})
-a^{2}(a\omega-m)^{2}(2r_{-}-r_{+}-r^{d}_{+}),
\end{equation}
\begin{equation}
C=\frac{eQ}{1+\alpha}\Big(\omega r^{2}_{-}(r_{+}r_{-}+r^{2}_{-}-3r_{+}r^{d}_{+}+r_{-}r^{d}_{-})
-a(a\omega-m)(r_{+}r_{-}-3r^{2}_{-} 
+r_{+}r^{d}_{+}+r_{-}r^{d}_{+})\Big) 
+ \big(\frac{eQ}{1+\alpha}\big)^{2} r_{-}(-r^{2}_{-}
+r_{+}r^{d}_{+}),
\end{equation}
\begin{equation}
D=\frac{2isa^{2}(1+\alpha)}{\alpha} \Big(\frac{\omega(r_{-}r^{d}_{-}+a^{2})-am -\frac{eQ}{1+\alpha}
\frac{r_{-}+r^{d}_{-}}{2}}{(r^{d}_{+} -r^{d}_{-})(r_{+}-r_{-})(r_{-}-r^{d}_{-}) }\Big)
  +(s+1)(2s+1)\Big(\frac{2(r^{d}_{-})^2}{(r_{+}-r_{-})(r^{d}_{+}-r^{d}_{-})}-z_{inf} \Big),
\end{equation}
\begin{equation}
E=-2B_{1}(z_{r} B_{2}+B_{3}) -\frac{a^{2}}{\alpha(r_{+}-r_{-})(r^{d}_{+}-r^{d}_{-})}
\big(-\lambda-2iesQ+2s(1-\alpha)\big)
-(s+1)\big((1+z_{r})B_{1}+z_{r} B_{2}+B_{3}\big).
\end{equation}
$T$ is the discriminant of $\Delta=0$,
\begin{equation}
T=(r_{+}-r_{-})^{2} (r_{+}-r^{d}_{-})^{2}(r_{+}-r^{d}_{+})^{2} (r_{-}-r^{d}_{-})^{2}
 (r_{-}-r^{d}_{+})^{2}(r^{d}_{+}-r^{d}_{+})^{2}
=\frac{16a^{10}}{\alpha}(F+H),
\end{equation}
where
\begin{equation}
F= (1-\alpha)^{3}\big(M^{2}-(1-\alpha)(a^{2}+Q^{2})\big),
\end{equation}
\begin{equation} 
H=\frac{\alpha}{a^{2}}\big(-27M^{4}+36(1-\alpha)M^{2}(a^{2}+Q^{2}
-8(1-\alpha)^{2}(a^{2}+Q^{2})^{2}\big)- \frac{16\alpha^{2}}{a^{4}} (a^{2}+Q^{2})^{3}.
\end{equation}
After the differential equation is reduced to an equation with only four regular singular points, one can write the solutions in terms of 
General Heun functions. We find that aside from the terms raised to given powers multiplying it, the solution of Eq. (10) is 
\begin{equation}
g(z)= H_{G} ( z_{r}, -v;\sigma_{+}, \sigma_{-}; 2B_{1}+s+1,2B_{2}+s+1;z)
\end{equation}
in the standard given in \cite{Arscott}.

We may also look for polynomial solutions for our Heun equation for the argument $u$. Using similar words as in \cite{Hortacsu1} we state: In \cite{Vieira,vie}, the boundary conditions to have resonant frequencies are given as to have the radial solution to be finite at the horizon, in our case at the origin, and well behaved at asymptotic 
infinity, which is at the cosmological horizon for the dS or at asymptotic infinity for the AdS cases. Vieira et al. state that to satisfy the condition at
asymptotic infinity one needs polynomial solutions. The first requirement for this is to have either $\sigma_+$ or $\sigma_-$ equal to $ -n$ \cite{ars}, i.e. "These frequencies are the proper modes at which a black hole freely oscillates when excited by a perturbation... Since they are damped by the emission of gravitational waves, the corresponding eigenvalues are complex \cite{sak}". $ n $ is the rank of the polynomial. Then our solution for the variable $ z$ is given by
\begin{equation}
R(z)=z^{B_1}(z-1)^{B_2}(z-z_{r})^{B_3} (z-z_{inf})^{2s+1}g(z).
\end{equation}

There is a second necessary criterion, which is the vanishing of a determinant given in \cite{cif,kar}, which fixed the value of the parameter $\lambda$ is fixed.

We have to note that, when $\sigma_+ = -n$, we get an complex value for $\omega$. Furthermore, the same $\omega$ is used in the time dependence of the solution as 
$e^{i \omega t}$. We have both a propagating and a decaying wave, since $i \omega$ has both real and imaginary parts.

\section{The reflection coefficients}
\label{sec:3}

Here we use the information we obtained from \cite{Suzuki} and try to calculate the possible scattering for waves coming from the de Sitter horizon at the outer horizon. 
We first calculate the two solutions at $r^{d}_{-}$, then use the connection formula given in \cite{Dekar} to write the solution at the outer horizon in terms the wave coming from and reflected to the inner de Sitter horizon. Unfortunately, this formula \cite{Dekar} works only between two finite points. That is why we first use a transformation of the independent variable $z$ in Eq. (3) as
\begin{equation}
    t=\frac{1-z}{z-z_{r}}.
\end{equation}
Then the new variable will be zero at the outer horizon, and $-1$ at the inner de Sitter horizon. Note that this transformation yields the following equation in terms of the new independent variable $t$ as
\begin{equation}
\frac{d^{2}R}{dt^{2}}
+( \frac {1-(\sigma_{+}+\sigma_{-})}{t+1} +\frac{2B_{2}+s+1}{t} +\frac{2B_{1}+s+1}{t+\frac{1}{z_{r}}} )\frac{dR}{dt}
 +\big((\sigma_{+}\sigma_{-})(\frac{1}{(t+1)^{2}})- \frac{M}{t(t+1)(z_{r}t+1)}\big)R=0.
\end{equation}
This equation is not of the Heun form. To put it to the Heun form, we make a s-homotopic transformation
\begin{equation} 
R= (t+1)^{\kappa} S_{1}(t)
\end{equation}
and find the first solution as $\kappa=\sigma_{+}$, and the second as $\kappa= \sigma_{-}$. We choose the first solution, which yields the 
differential equation
\begin{eqnarray}
&&\frac{d^{2}S_{1}}{dt^{2}}
+( \frac {1+\sigma_{+}-\sigma_{-}}{t+1} +\frac{2B_{2}+s+1}{t} +\frac{2B_{1}+s+1}{(t+\frac{1}{z_{r}})})\frac{dS_{1}}{dt} {}\nonumber \\&&
+\big(\frac{\sigma_{+}(\sigma_{+}-s-B_{3})}{(t+1)(t+\frac{1}{z_{r}})}
-\frac{v-\sigma_{+}(s-B_{2}+1)}{t(t+1)(tz_{r}+1)}\big)S_{1}=0.
\end{eqnarray}
This equation is not in the standard Heun form.
Recalling that here $t$ is between $-1$ to $zero$, we define $u$ as the absolute value of $t$. Then the solution is
\begin{eqnarray}
S_{1}=H_{G}(\frac{1}{z_{r}},-v+\sigma_{+}(s-B_{2}+1);\sigma_{+},\sigma_{+}-s-B_{3}; 
2B_{2}+s+1,1+\sigma_{+}-\sigma_{-};u).
\end{eqnarray}
In the standard form \cite{Arscott}, the parameters given as 
$H_{G}(a,q;\alpha,\beta; \gamma,\delta;z)$ for the differential equation
\begin{equation}
\frac{d^{2}H_G (x)}{d x^{2}} +\big( \frac{\delta}{x-1} + \frac{\epsilon}{x-a}+ \frac{{\gamma}}{{x}}
\big) {\frac{{dH_G (x) }}{{d x }}}
+\big(\frac{{\alpha \beta x-Q}}{{x(x-1)(x-a)}} \big) H_G(x) = 0.
\end{equation}
In \cite{Dekar}, the authors, in their equation (A.15), for a finite interval, give  the  necessary formulae, to write the $H_{G}(z)$, solution 
expanded in terms of $z$, in terms of two solutions of the equation, expanded
around $1-z$. This is a formula to write
\begin{equation}
 y_{1}(z)= C_{1} y_{3}(1-z)+C_{2}(1-z)^{\kappa} y_{4}(1-z).
\end{equation}
This formula, for general Heun functions, reads
\begin{eqnarray}
&&H_{G}( a,Q;\alpha,\beta; \gamma,\delta;z)=C_{1}H_{G}(1-a, -Q-\alpha\beta ; \alpha,\beta; 
\alpha+\beta-\gamma-\delta,\delta;1-z) 
\nonumber \\&&
+C_{2}(1-z)^{\gamma+\delta-\alpha-\beta} 
 H_{G}(1-a,Q*;
\gamma+\delta-\alpha, \delta+\gamma-\beta; 
\gamma+\delta-\alpha-\beta,\delta;1-z).
\end{eqnarray}
Here 
\begin{equation}
C_{1}=H_{G}(a,q;\alpha,\beta; \gamma,\delta;1),
\end{equation}
\begin{equation}
C_{2}=H_{G}(a,-Q-a\gamma[\gamma+\delta-\alpha-\beta];
\gamma+\delta-\alpha,\gamma+\delta-\beta;\gamma,\delta;1),
\end{equation}
\begin{equation}
Q*=-Q-\alpha\beta-[\gamma+\delta-\alpha-\beta][\gamma+\delta-b\gamma].
\end{equation}

In our example, we have
\begin{equation}
y_{1}= H_G(\frac{1}{z_{r}},Q_{1};
\sigma_{+},\sigma_{+}-B_{3}-s;
1+\sigma_{+}-\sigma_{-},2B_{2}+s+1;u),
\end{equation}
\begin{equation}
Q_{1}=  -\frac{v-\sigma_{+}(2B_{2}+s+1)}{z_{r}},
\end{equation}
\begin{equation}
y_{3}=H_{G}(1-\frac{1}{z_{r}},
Q_{3};\sigma_{+},\sigma_{+}-B_{3};
1+\sigma_{+}-\sigma_{-},2B_{2}+s+1;1-u),   
\end{equation}
\begin{equation}
Q_{3} =\frac{v-\sigma_{+}(2B_{2}+s+1}{z_{r}}+\sigma_{+}(\sigma_{+}-s-B_{3}),   
\end{equation}
\begin{equation}
C_1=H_{G}(1-\frac{1}{z_{r}}, -\frac{v-\sigma_{+}(2B_{2}+s+1)}{z_{r}};
\sigma_{+},\sigma_{+}-2B_{3}-s; 2B_{1}+s+1,2B_{2}+s+1;1).
\end{equation}
In terms of the parameters giving in the original equation, one writes
\begin{equation}
\kappa=-\sigma_{+}+\sigma_{-}=  
-i( \frac{2(1+\alpha)a^{2} (\omega((r^{d }_{-})^2 +  a^{2})-am-\frac{e Qr^{d}_{-}}{1+\alpha})}{\alpha(r^{d}_{+}- r_{+})(r^{d}_{+}-r_{+})(r_{-}-
r_{+})}-is),
\end{equation}
\begin{equation}
y_{4}
=H_{G}(1-\frac{1}{z_{r}},Q_{4} ;\sigma_{-},\sigma_{-}-2B_{3}-s;
1+\sigma_{-}-\sigma_{+},2B_{2}+s+1;1-u),
\end{equation}
\begin{equation}
Q_{4}= -\frac{v-\sigma_{-}(2B_{2}+s+1}{z_{r}}+\sigma_{-}(\sigma_{+}-s-B_{3}), 
\end{equation}
\begin{equation}
C_{2}=H_{G}(1-\frac{1}{z_{r}}, -\frac{v-\sigma_{-}(2B_{2}+s+1)}{z_{r}};
\sigma_{-},\sigma_{-}-2B_{3}-s;2B_{1}+s+1,2B_{2}+s+1;1).
\end{equation}    
Then the wave goes as, up to a decaying power,
\begin{equation}
e^{i\kappa ln (1-u)} y_{3}+ Re^{-i\kappa ln(1-u)} y_{4}.
\end{equation}
The reflection coefficient R is given by
\begin{equation}
R=|\frac{M}
 {N}|^{2}   
\end{equation}
Here $M$ and $N$ are given by
\begin{equation}
M= H_{G}(1-\frac{1}{z_{r}}, -\frac{v-\sigma_{-}(2B_{2}+s+1)}{z_{r}};
\sigma_{-},\sigma_{-}-2B_{3}-s;2B_{1}+s+1,2B_{2}+s+1;1),
\end {equation}
\begin{equation}
N=H_{G}(1-\frac{1}{z_{r}}, -\frac{v-\sigma_{+}(2B_{2}+s+1)}{z_{r}};
\sigma_{+},\sigma_{+}-2B_{3}-s;2B_{1}+s+1,2B_{2}+s+1;1).
\end {equation}
Another application will be the reflection of waves coming from the outer horizon at the inner horizon. In \cite{Suzuki}, it is stated that, the transformation they used brought the inner horizon to $z=0$, and the outer horizon to $z=1$. To get the reflection term, we need one solution 
around $z=0$, and two solutions in terms of the  the transformed independent variable at $z=1$. For a solution around $z$, we use the differential 
equation, our Eq. (10), given in \cite{Suzuki}. The total solution is given in our Eq. (6) \cite{Suzuki}. The factors that multiply the Heun 
function are common to all three solutions. The Heun part of the solution is given by
\begin{equation}
  Y_{1}= H_{G}(z_{r}, -v; \sigma_{+}, \sigma_{-};2B_{1}+s+1,2B_{2}+s+1; z).
\end{equation}
We, then, translate the variable to $1-z$ with the new Heun solution
\begin{equation}
 Y_{3}= H_{G}(1-z_{r}, v+\sigma_{+}\sigma_{-}; \sigma_{+}, \sigma_{-};
2B_{2}+s+1,2B_{1}+s+1;1-z).
\end{equation}
We find the second solution by multiplying $Y_{1}$ by $(1-z)^{\psi}$ and looking for the proper value of ${\psi}$ to get a Heun type solution. 
We find $\psi$ equal to $-(2B_{2}+s)$ to give us
\begin{equation}
Y_{4}=(1-z)^{-(2B_{2}+s)}  H_{G}(1-z_{r}, Q_{4};
\sigma_{-}-(2B_{2}+s), \sigma_{+}-(2B_{2}+s);
1-(2B_{2}+s), 1+2B_{1}+s ;1-z) 
\end{equation}
\begin{equation}
 Q_{4}=v +\sigma_{+}\sigma_{-}+(2B_{2}+s)((2B_{1}+s)(1-z_{r})+2B_{3}+s+1).
\end{equation}

In our Eq. (12) giving $B_2$, we choose the plus sign. Then $\psi$ is an imaginary quantity, aside from $-s$, giving a decaying solution, as it 
approaches the inner horizon. After multiplying both sides of the equation
\begin{equation}
Y_{1}= D_{3} Y_{3}+ (1-z)^{-(2B_{2}+s)}D_{4} Y_{4} ,
\end{equation}
by $(1-z)^{(B_{2}+\frac{s}{2})}$, we obtain an equation of the form
\begin{equation}
 (1-z)^{(B_{2}+\frac{s}{2})} Y_{1}= (1-z)^{(B_{2}+\frac{s}{2})}D_{3} Y_{3}+ (1-z)^{-(B_{2}+\frac{s}{2})} D_{4} Y_{4}.  
\end{equation}
Up to an overall constant, this equation may be written as an incoming and outgoing waves, as a function of $1-z$ for $1>z>0$, equal to the wave 
evaluated at $z$ around zero.

For this case the reflection constant $R_{i}$ is given as
\begin{equation}
R_{i}=|\frac{P}{H_{G}(z_{r}, -v-;\sigma_{+};\sigma_{-};2B_{1}+s+1,2B_{3}+s+1;1)}|^{2},
\end{equation}

\begin{equation}
P=H_{G}(z_{r},Q_{5} ;\sigma_{-}-(2B_{1}+s),\sigma_{+}-(2B_{1}+s);
2B_{1}+s+1,2B_{3}+s+1;1),
\end{equation}
\begin{eqnarray}
Q_{5}=-v-z_{r}(2B_{1}+s+1)(2B_{1}+s).
\end{eqnarray}

Note that we have two formal results, our Equations (44) and (53). They are formal solutions, since we do not know if our Heun solution is 
convergent at $u$ and $z$ equal to unity. We can use the analysis as done by Leaver \cite{Leaver}, and conclude that to have a convergent  
Heun function, we need, in Eq. (51)  $z_r>1$, and $z_r<1$ in Eq. (43). Since these two results are incompatible, we choose the first case, 
giving in our equation (43). Our analysis is correct only for reflection for waves coming from the de Sitter horizon and scatter at the outer horizon.

\section{Conclusion}
We used the wave equation, obtained, reduced to a manageable form by the authors in \cite{Suzuki}, and solved for different values of the 
independent variable in terms of General Heun functions, for the regions between the de Sitter and the outer horizons, and between the outer and 
inner horizons. They \cite{Suzuki,Suzuki1} used infinite series expansions of the Heun function. We used the Heun functions directly. We tried to 
calculate the reflection coefficients for waves for this two regions formally. Unfortunately, we could not get convergent solutions at two regular 
singular points for both of these regions, by putting constraints on the parameters in the wave equations \cite{Leaver}. We chose to use the  
constraint, $z_r<1$. Note that our  Eq. (43) is a consistent equation, and our Eq. (51) may be inconsistent, since the constant $C_{i}, D{i}$, we used, 
may not be finite. We may, therefore, dismiss the latter case. The similar problem may be studied in similar metrics \cite{Finnian,Frolov}.
\section* {Acknowledgement}
We thank Prof. Tolga Birkandan for collaboration in the early part of this work, and for given me our references [6] and [10]. I thank Prof. N. 
Ghazanfari for giving me my key reference [20]. I thank Prof. Reyhan Kaya for technical assistance. This work is morally supported by Science 
Academy, Istanbul.

\end{document}